\documentclass[%
preprint,
 amsmath,amssymb,
 aps
]{revtex4-2}

\usepackage[english]{babel}
\usepackage{colortbl}
\usepackage{dcolumn}
\usepackage{bm}
\usepackage{multirow}
\usepackage{supertabular}
\usepackage{inputenc}
\usepackage{makecell}

\newcommand{\expone}{\textbf{1}$^{\mathrm{exp}}$}
\newcommand{\exptwo}{\textbf{2}$^{\mathrm{exp}}$}

\newcommand{\modelone}{\textbf{1}$^{\mathrm{model}}$}
\newcommand{\modeltwo}{\textbf{2}$^{\mathrm{model}}$}

\usepackage{graphicx}
\usepackage{hyperref}

\begin{document}

\preprint{}

\title{Structural Distortions Control Scaling of Exciton Binding Energies in Two-Dimensional Ag/Bi Double Perovskites}

\author{Pierre Lechifflart}
 \affiliation{MESA+ Institute for Nanotechnology, University of Twente, 7500 AE Enschede, The Netherlands}
  \email{p.lechifflart@utwente.nl}

\author{Raisa-Ioana Biega}%
\affiliation{MESA+ Institute for Nanotechnology, University of Twente, 7500 AE Enschede, The Netherlands}%

\author{Linn Leppert}
 \affiliation{MESA+ Institute for Nanotechnology, University of Twente, 7500 AE Enschede, The Netherlands}%
 \email{l.leppert@utwente.nl}

\date{\today}


\begin{abstract}
Three-dimensional metal halide double perovskites such as Cs$_2$AgBiBr$_6$ exhibit pronounced excitonic effects due to their anisotropic electronic structure and chemical localization effects. Their two-dimensional derivatives, formed by inserting organic spacer molecules between perovskite layers, were expected to follow well-established trends seen in Pb-based 2D perovskites, namely, increasing exciton binding energies with decreasing layer thickness due to enhanced quantum and dielectric confinement. However, recent experimental and computational studies have revealed anomalous behavior in Ag/Bi-based 2D perovskites, where this trend is reversed.
Using ab initio many-body perturbation theory within the $GW$ and Bethe-Salpeter Equation frameworks, we resolve this puzzle by systematically comparing experimental structures with idealized models designed to isolate the effects of octahedral distortions, interlayer separation, and stacking. We find that structural distortions, driven by directional Ag d orbital bonding, govern the momentum-space origin and character of the exciton, and are the primary cause of the observed non-monotonic trends. Furthermore, we explore how interlayer distance and stacking influence band gaps and exciton binding energies, showing that, despite different chemistry, the underlying confinement physics mirrors that of Pb-based 2D perovskites. Our results establish design principles for tuning excitonic properties in this broader class of layered, lead-free materials.
\end{abstract}

\maketitle


Halide double perovskites with chemical formula A$_2$BB'X$_6$, where A is a monovalent cation, B and B' are metal cations with alternating oxidation states of +I and +III, and X is a halide, constitute a versatile family of semiconductors with a wide range of tunable optoelectronic properties.\cite{Wolf2021} As lead-free alternatives to the quintessential metal-halide perovskite CH$_3$NH$_3$PbI$_3$, they have been explored for a wide variety of applications from photovoltaics \cite{sirtl2D3DHybrid2022, zhangHydrogenatedCs2AgBiBr6Significantly2022a} and photocatalysis~\cite{muscarellaHalideDoublePerovskiteSemiconductors2022a} to detectors \cite{panCs2AgBiBr6SinglecrystalXray2017, Steele2018}. Ag-pnictogen double perovskites, Cs$_2$AgBX$_6$ with B=Bi$^{3+}$, Sb$^{3+}$ and X=Br$^-$, Cl$^-$, and in particular the double perovskite Cs$_2$AgBiBr$_6$, have been widely studied due to their thermodynamic stability and interesting photophysical properties, including an indirect band gap in the visible range \cite{McClure:cm2016, Volonakis:jpcl2016}, good charge-carrier mobilities~\cite{Slavney:jacs2016}, and robust stability~\cite{Schade2019}. Furthermore, recent studies have shown that halide double perovskites can host excitons - correlated electron-hole pairs - with binding energies ranging from tens of meV to $\sim$2\,eV, exhibiting either hydrogenic Wannier-Mott or more strongly localized Frenkel-like character depending on composition~\cite{cuccoFineStructureExcitons2023, biegaChemicalMappingExcitons2023, leppertExcitonsMetalhalidePerovskites2024}. The optical absorption spectrum of Cs$_2$AgBiBr$_6$ is characterized by a strongly localized resonant exciton~\cite{Palummo:acsEL2020} ill-described by the Wannier-Mott model due to the chemical heterogeneity of this material that is reflected by a highly anisotropic electronic structure and dielectric screening~\cite{Biega:jpcl2021}. 

Layered organic-inorganic halide perovskites are quasi-twodimensional (2D) derivatives of the three-dimensional (3D) halide perovskites, which consist of alternating layers of corner-sharing metal-halide octahedra and organic molecules, which can be assembled in myriad of different structures, depending on the composition, structure, and bonding interactions of the organic and inorganic sublattices~\cite{Smith:armr2018}. Since the electronic coupling between metal-halide octahedra is disrupted by the organic sublattice in these materials, their photophysical properties are dictated by quantum and dielectric confinement effects~\cite{blanconSemiconductorPhysicsOrganic2020} and thus vary strongly with interlayer distance (which can be controlled through the length of the molecular cation constituting the organic sublattice) \cite{leveilleeTuningElectronicStructure2019}, the dielectric constant of the organic sublattice~\cite{Smith2017a, Filip2022}, the stacking pattern of adjacent inorganic layers~\cite{chenTunableInterlayerDelocalization2023}, and most notably, the thickness of the inorganic sublattice \cite{Blancon2018, Cho2019}, typically denoted by $n$, the number of metal-halide octahedra in the out-of-plane direction. For example, in lead-based layered perovskites, A$_2$A'$_{n-1}$Pb$_n$X$_{3n+1}$, where A is, e.g., butylammonium (BA) or phenylethylammonium (PEA), pronounced changes in the optical band gap and exciton binding energies have been reported based on experiments and computational modelling~\cite{Tsai2016, cao_2d_2015, chakraborty_quantum_2022}, supported by microscopic theories of quantum and dielectric confinement effects in these materials~\cite{Tanaka2005, Blancon:natcomm2018, Cho2019}. These studies show that the exciton binding energy decreases significantly with increasing $n$, approaching the exciton binding energy of bulk ABX$_3$ with an exponential dependence on $n$.

Layered derivatives of Cs$_2$AgBiBr$_6$ were first reported in 2018 as (BA)$_4$AgBiBr$_8$ (corresponding to $n=1$ and denoted by \textbf{1}$^{\mathrm{exp}}$ in the following) and (BA)$_2$CsAgBiBr$_{7}$ ($n=2$ and denoted by \textbf{2}$^{\mathrm{exp}}$)~\cite{Connor2018}. First-principles density functional theory (DFT) calculations of these materials showed that dimensional reduction leads to a transition from an indirect ($n \geq 2$) to a direct band gap ($n=1$) due to the two-dimensional symmetry of the lattice, the effect of spin-orbit coupling and structural distortions \cite{Connor2018, connorUnderstandingEvolutionDouble2023}. This observation is reminiscent of the indirect -- direct transition observed in some transition metal dichalcogenides upon exfoliation to the monolayer limit \cite{splendiani2010emerging}. In Ref.~\citenum{Connor2018}, \expone{} and \exptwo{} were shown to both exhibit sharp particle-like features in their optical absorption spectra, at similar energies. Furthermore, polycrystalline powders of \expone{} and \exptwo{} had almost identical (yellow) color. These findings suggested the puzzling observation that confinement effects are less pronounced in these systems compared to their Pb-based counterparts. 

Similar conclusions were drawn by Pantaler \textit{et al.} for (PEA)$_4$AgBiBr$_8$ ($n=1$) and (PEA)$_2$CsAgBiBr$_{7}$ ($n=2$)~\cite{pantalerRevealingWeakDimensional2022}, Ref.~\citenum{pantalerRevealingWeakDimensional2022} based on experimentally obtained exciton binding energies and rationalized with first-principles calculations using the Bethe-Salpeter Equation (BSE) with electronic band energies from DFT~\cite{pantalerRevealingWeakDimensional2022}. Full $GW$+BSE calculations of experimental structures of \expone{} and \exptwo{} were reported by Palummmo \textit{et al.} who predicted exciton binding energies of 700\,meV for \expone{} and 970\,meV for \exptwo{}\cite{Palummo:apl2021}, and similar absorption onsets, in line with experimental results~\cite{Connor:jacs2018, Pantaler:jacsau2022}, but without providing an explanation for these counterintuitive trends.

Here, we show that confinement effects in 2D Ag/Bi perovskites are similar to those in Pb-based 2D perovskites and provide an explanation for their anomalous exciton binding energy trends. In stark contrast to their Pb-based counterparts, excitonic properties of 2D Ag/Bi perovskites are highly sensitive to structural distortions, which differ between systems with $n=1$ and $n=2$. As a consequence, excitons in $n=1$ and $n=2$ Ag/Bi double perovskites arise from different reciprocal space points and have different character. We demonstrate this by systematically investigating the electronic bandstructures and excitonic effects using first-principles calculations based on DFT, and the $GW$ and BSE approaches, and use model systems to disentangle the effects of the molecular A site, structural distortions, interlayer distance, and interlayer stacking. Furthermore, our calculations demonstrate the rich diversity of electronic and excitonic properties that can be achieved by varying interlayer distance and stacking motifs, for example with different organic sublattices.

\begin{figure}[htb]
    \centering
    \includegraphics[width=\linewidth]{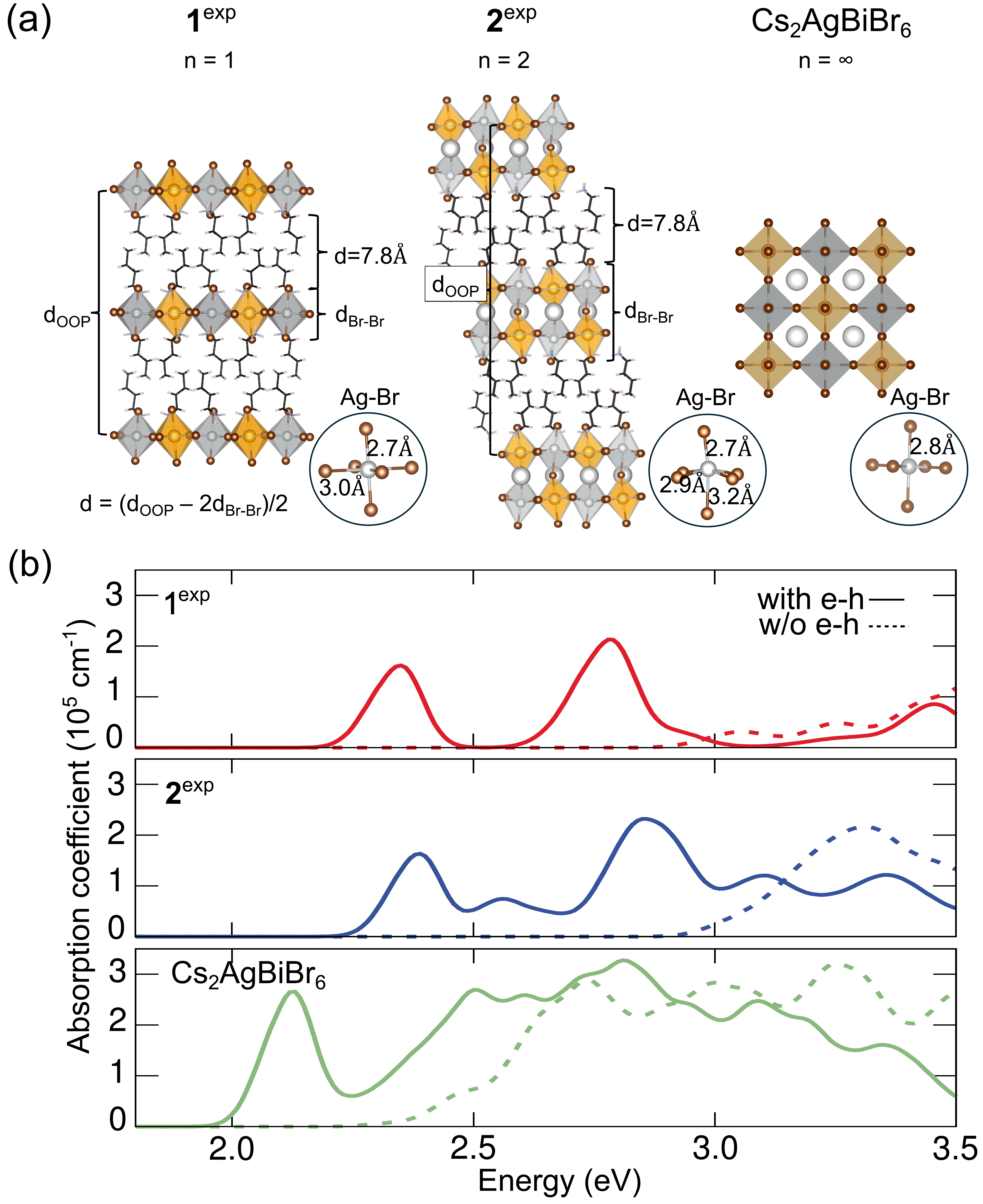}
    \caption{(a) Crystal structures of \expone{}, \exptwo{} and Cs$_2$AgBiBr$_6$, where Ag is represented by silver, Bi by orange, Br by brown, and Cs by white balls, and N by blue, C by black, and H by pink sticks. Bondlength variations in Ag-Br octahedra are highlighted separately. (b) $G_0W_0$@PBE+SOC + BSE (with e-h, solid lines) and independent particle (w/o e-h, dashed lined) absorption spectra of \expone{} (red), \exptwo{} (blue), and Cs$_2$AgBiBr$_6$ (green).}
    \label{fig:structures-absorption}
\end{figure}
We start by calculating the electronic structure, optical absorption spectra, and exciton binding energies of \expone{} and \exptwo{}. The electronic properties were first calculated with DFT using the Perdew-Burke-Ernzerhof (PBE) functional \cite{Perdew:prl1996} as implemented in \texttt{QuantumESPRESSO} \cite{Giannozzi:jpCM2009,Giannozzi:jpCM2017, Giannozzi:jcp2020}, including spin-orbit coupling self-consistently. For these calculations, we used experimental structures determined using X-ray crystallography at 100\,K from Ref.~\citenum{Connor2018}, as shown in Figure~\ref{fig:structures-absorption}a. Both \expone{} and \exptwo{} are 2D perovskites of the Ruddlesden-Popper form, i.e., with a shift along half the in-plane diagonal between adjacent perovskite layers. As noted before, the Ag-Br octahedra in \expone{} and \exptwo{} are strongly distorted as compared to those in cubic Cs$_2$AgBiBr$_6$ \cite{Connor2018}. Terminal Ag-Br bonds, i.e., those that are in the vicinity of the organic sublattice, are shorter, whereas both equatorial and bridging Ag-Br bonds are elongated as compared to the Ag-Br bonds in Cs$_2$AgBiBr$_6$.

We define the interlayer distance in these and all other structures mentioned in the following as $d = (d_{\mathrm{OOP}} - 2d_{\mathrm{Br-Br}})/2$, where $d_{\mathrm{OOP}}$ is the out-of-plane distance between equivalent layers, and the width of the inorganic layer is defined as the average distance between terminal Br atoms $d_{\mathrm{Br-Br}}$, as indicated in Figure~\ref{fig:structures-absorption}a. Using these definitions, the interlayer distance of both \expone{} and \exptwo{} is $d \sim$ 7.8\,\AA.

We then used the $GW$+BSE approach to calculate the electronic band gaps and linear optical absorption spectra of Cs$_2$AgBiBr$_6$, \expone{}, and \exptwo{}, as shown in Figure~\ref{fig:structures-absorption}b. These calculations were performed with the \texttt{BerkeleyGW} package, using PBE eigenvalues and eigenfunctions, including spin-orbit coupling (SOC), to construct the $G_0W_0$ self-energy, followed by solution of the BSE. We find that \expone{} has a direct gap at $\Gamma$, whereas \exptwo{} and Cs$_2$AgBiBr$_6$ are indirect gap semiconductors, in agreement with previous calculations and models~\cite{Connor2018, Palummo:apl2021, Pantaler:jacsau2022, connorUnderstandingEvolutionDouble2023}. G$_0$W$_0$@PBE+SOC direct band gap energies are reported in Table~\ref{tab:bandgaps}. Figure~\ref{fig:structures-absorption}b shows that while Cs$_2$AgBiBr$_6$ exhibits one excitonic peak arising from the X point \cite{Palummo2020, Biega2021a}, \expone{} has two distinct excitonic peaks below the onset of the independent-particle absorption, separated by approximately 0.5\,eV. These correspond to two excitonic states, arising from transitions at $\Gamma$ (lower energy) and A (higher energy), respectively.

\begin{figure}[htb]
    \centering
    \includegraphics[width=\linewidth]{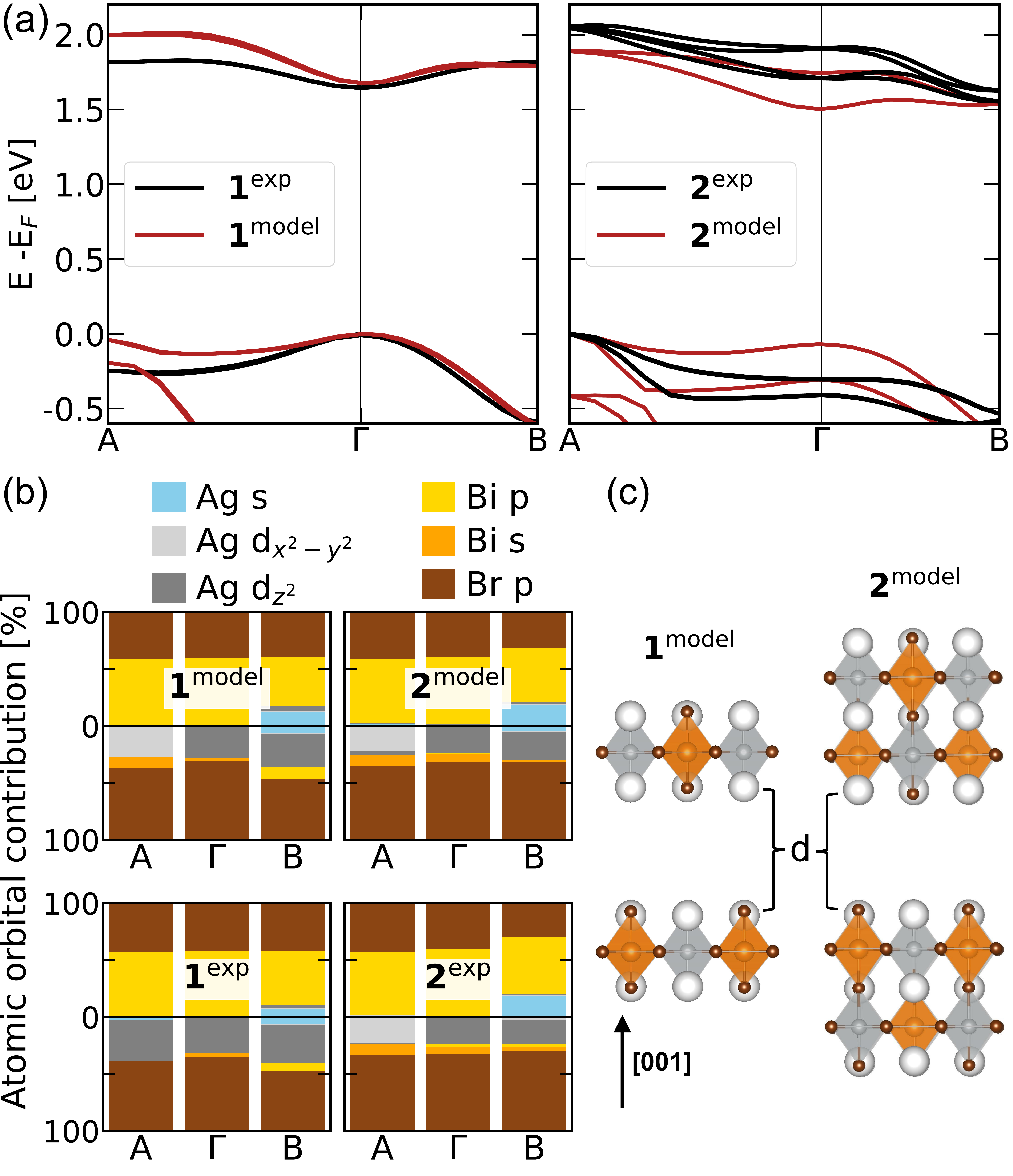}
    \caption{(a) left : DFT-PBE+SOC electronic bandstructures of \expone{} (brown) and \modelone{} (black); right : \exptwo{} (brown) and \modeltwo{} (black). (b) Atomic orbital contributions within DFT-PBE+SOC at three high-symmetry points,  for the top valence (bottom half) and bottom conduction band (top half), for \modelone{} and \modeltwo{} (top row), and \expone{} and \exptwo{} (bottom row). (c)  Crystal structures of RP model systems with $n=1$ and $n=2$, with d$_{\mathrm{Br-Br}}$ = 5.6\,\AA{} in \modelone{} and d$_{\mathrm{Br-Br}}$ = 11.2\,\AA{} in \modeltwo{}.}
    \label{fig:exp-vs-model}
\end{figure}

\exptwo{} also shows two distinct peaks coming from two different excitonic states, separated by approximately 0.6\,eV. This time, the lower-energy peak comes from transitions at A, while the higher-energy one comes from transitions at $\Gamma$. More evidence for this inversion is shown in Table~\ref{tab:bandgaps} and Figure~\ref{fig:exp-vs-model} and discussed later. The onset of absorption of \expone{} and \exptwo{} is at similar energies, in agreement with experimental observations~\cite{Connor2018}. However, we note that our $G_0W_0$@PBE+BSE calculations underestimate the energy of the experimental peaks by $\sim$0.5 - 0.8\,eV, which is typical for $G_0W_0$ calculations of halide perovskite band gaps based on semilocal DFT \cite{Filip2014, Leppert2019l}. Our absorption spectra are in good agreement with computational results by Palummo \emph{et al.}, with differences arising from the use of ev$GW$ and different numerical parameters in Ref~\citenum{Palummo2021}. All relevant computational parameters and convergence tests, as well as a discussion of the differences with Ref.~\citenum{Palummo2021} can be found in the Supplementary Information.

\begin{table*}
\centering
\begin{tabular}{l|ccc|c|cc|cc}
System     & \multicolumn{3}{c}{$G_0W_0$@PBE direct gaps (eV)} \vline  & $\epsilon_{\infty}$ & \multicolumn{2}{c}{\makecell{Exciton \\ binding energy (meV)}} \vline & \multicolumn{2}{c}{Exciton}\\
             & A    & $\Gamma$ & B        &         & dark    & bright  & origin    & extent (\AA)         \\ \hline \hline
\expone      & 3.47 & 2.96     & 3.97     & 3.03    & 736     & 654     & $\Gamma$  & 8.14   \\
\exptwo      & 3.00 & 3.10     & 3.27     & 3.63    & 893     & 734     & A         & 8.24   \\
\modelone    & 3.06 & 2.78     & 3.80     & 3.22    & 664     & 608     & $\Gamma$  & 14.6   \\
\modeltwo    & 3.02 & 2.76     & 3.64     & 3.77    & 592     & 513     & $\Gamma$  & 11.0   \\ \hline
Cs$_2$AgBiBr$_6$ & 2.16 (X) & 2.23 & 2.96 (L)  & 5.41  & 260  & 181     & X        &  7.52   
\end{tabular}
\caption{Energy differences between conduction and valence band within $G_0W_0$@PBE+SOC, static dielectric constant as computed within the random phase approximation (left). The high-symmetry points A and B of the 2D Brillouin Zone are equivalent to the X and L points of the face-centered cubic lattice, respectively~\cite{Connor2018}.
Binding energy of the first dark and first bright excitons with their origin in the BZ (right). Values for 3D Cs$_2$AgBiBr$_6$ are extracted from Ref.~\citenum{biega2024mixing}. Spatial extent of lowest-energy exciton.}
\label{tab:bandgaps}
\end{table*}

To understand the counterintuitive trends in the band gaps and exciton binding energies of \expone{} and \exptwo{}, we constructed several model systems (see Supplementary Information). First, we replaced the organic cations in the experimental structures with Cs while preserving their structural distortions. Since Cs-derived states do not contribute near the band edges, the resulting band structures remain nearly identical to those of \expone{} and \exptwo{}. The reduced dielectric screening due to the absence of organic molecules increases exciton binding energies but preserves the qualitative trends observed in the experimental systems (Table S1).

Conversely, these trends change in the model structures \modelone{} and \modeltwo{}, where the organic cations are replaced by Cs and all octahedral distortions are removed. These structures feature uniform Ag–Br and Bi–Br bond lengths, identical to those in Cs$_2$AgBiBr$_6$ (Figure~\ref{fig:exp-vs-model}c). Their DFT-PBE+SOC band structures (Figure~\ref{fig:exp-vs-model}a) show nearly degenerate valence band maxima at $\Gamma$ and A, derived from Ag $d_{z^2}$ and $d_{x^2 - y^2}$ orbitals, and a direct band gap at $\Gamma$ with similar orbital character. Br $p$ contributions to the band edges are similar across all structures and omitted from further discussion (Figure~\ref{fig:exp-vs-model}b). In contrast, the shorter terminal Ag-Br and longer equatorial and bridging Ag-Br bonds of \exptwo{} lead to significant changes in the bandstructure, in particular a pronounced lowering of the valence band at $\Gamma$ and the conduction band at B, leading to an indirect band gap in \exptwo{} \cite{Connor2018, connorUnderstandingEvolutionDouble2023}. Thus our first main result is that octahedral distortions qualitatively change the band-edge positions, leading to a shift of the lowest-energy direct transition from $\Gamma$ in \expone{} to A in \exptwo{}.

We compute exciton binding energies for the lowest excitons in \modelone{} and \modeltwo{} using $G_0W_0$@PBE+SOC and BSE (Table~\ref{tab:bandgaps}). Our second key result is that the binding energies decrease with increasing layer thickness $n$, consistent with quantum confinement in 2D perovskites, though they remain larger than in Pb-based analogs. This effect is also evident in the absorption spectra, with a redshift of $\sim$0.25\,eV in \modelone{} relative to \modeltwo{} (Figure~S2). The spatial extent of the exciton, estimated from the average electron-hole separation \cite{Sharifzadeh2013}, is larger in the model structures, consistent with enhanced dielectric screening and reduced exciton binding (see Table~\ref{tab:bandgaps} and SI for details).

These results demonstrate that the anomalous trends in exciton binding energies in the experimental systems arise  from a change in their momentum-space origin: The lowest-energy exciton in \exptwo{} arises from a direct transition at the A point, whereas in \expone{} it derives from $\Gamma$. This shift is induced by octahedral distortions and results in a qualitatively different exciton due to symmetry and orbital-character differences at the A and $\Gamma$ points. We have previously observed a similar, but composition-driven, transition in 3D Cs$_2$AgBi(I$_x$Br$_{1-x}$)$_6$, where the lowest-energy direct transition shifts from L to $\Gamma$ with increasing iodine content due to changes in Ag $d$ orbital hybridization \cite{biega2024mixing}.

\begin{figure}[htb]
    \centering
    \includegraphics[width=\linewidth]{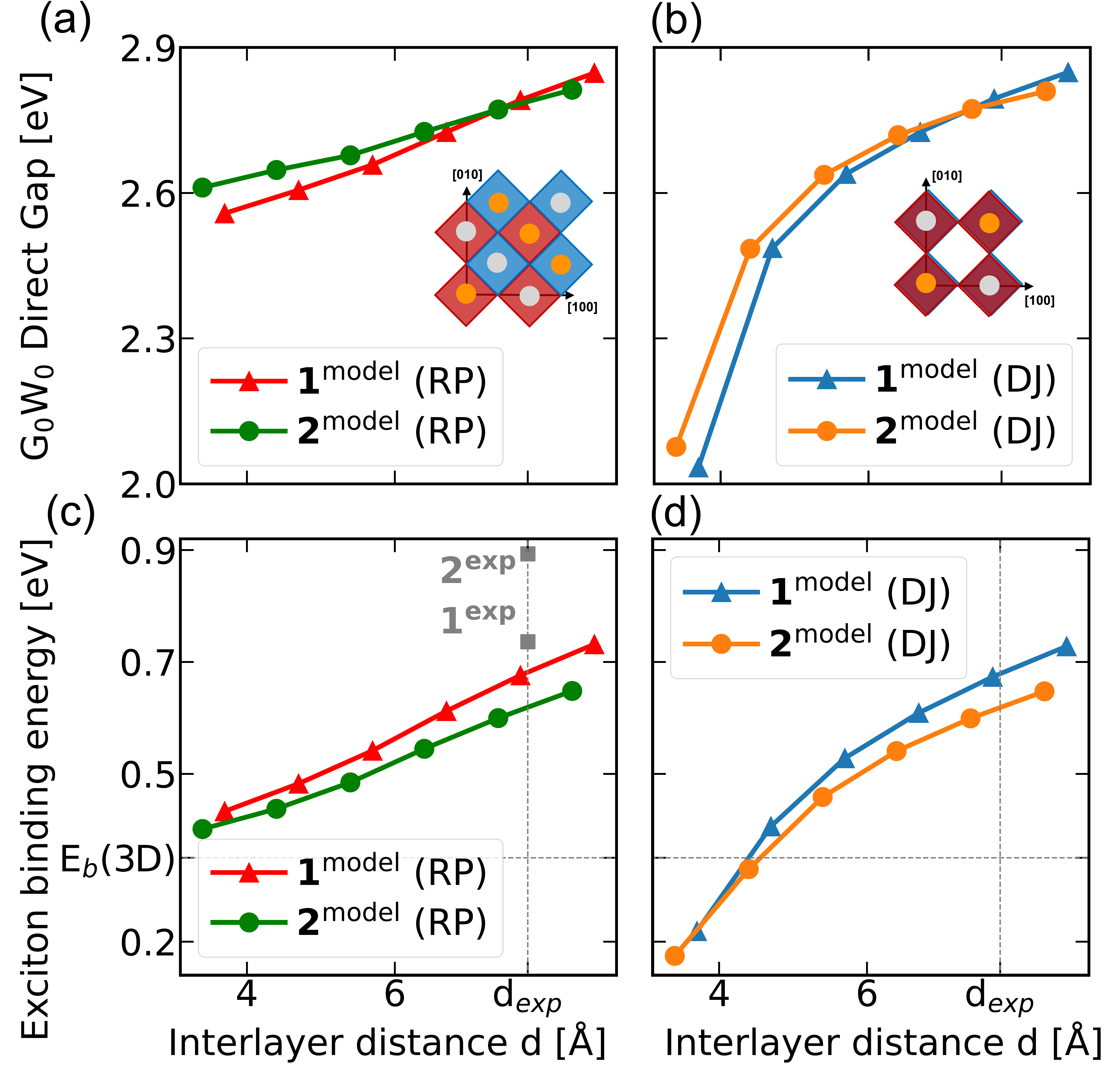}
    \caption{(a) G$_0$W$_0$ Direct gaps as a function of interlayer distance $d$ for the RP stacking pattern (inset) : Top view of RP stacking pattern in which adjacent layers, here pictured as red and blue octahedra, are shifted along half of the in-plane lattice diagonal. (b) and of the DJ stacking pattern (inset) Top view of the DJ stacking pattern in which adjacent layers are on top of one another. (c) Exciton binding energy as a function of interlayer distance $d$ for \modelone{} and \modeltwo{} in the RP stacking pattern and (d) in the DJ stacking pattern. Solid lines are a guide for the eye. }.
    \label{fig:eb-models}
\end{figure}
Having established the role of structural distortions, we now turn to the effect of interlayer distance and stacking pattern, i.e., Ruddlesden-Popper (RP) vs. Dion-Jacobson (DJ) stacking, on excitonic properties. In real 2D halide perovskites, interlayer spacing is typically controlled via organic cations, which also contribute significantly to dielectric screening \cite{Schmitz:cm2021, Filip:nl2022}. This means that any variation in interlayer distance using different molecules necessarily conflates geometric and dielectric effects. To disentangle these contributions, we use idealized model structures in which Cs replaces the organic spacers and all octahedral distortions are removed. In this case, the dielectric screening is significantly reduced and varies with interlayer distance, because the long-range Coulomb interaction is no longer screened.

We vary the interlayer spacing $d$ for \modelone{} and \modeltwo{} in both RP and DJ stacking (Figure~\ref{fig:exp-vs-model}c and d). The resulting direct band gaps and exciton binding energies are shown in Figure~\ref{fig:eb-models}. In all cases, we observe that increasing $d$ leads to stronger exciton binding and wider gaps, linearly in RP and as a fractional power law in DJ stacking. The lowest-energy direct transition remains at $\Gamma$, and \modelone{} consistently shows higher binding energies than \modeltwo{}, reinforcing that the anomalous trends in the experimental systems stem from octahedral distortions rather than confinement effects.

At large interlayer distances, the stacking motif has no effect on exciton binding, as interlayer interactions vanish and the excitons are confined to a single 2D layer. At smaller separations, electron and hole wavefunctions become more delocalized across layers, reducing both quantum and dielectric confinement and thereby lowering the exciton binding energy. In this regime, stacking becomes important: DJ stacking consistently yields lower binding energies than RP, regardless of layer thickness. This is due to the in-plane alignment of terminal Br atoms in DJ structures, which enhances Br $p$-orbital overlap as the layers approach. At very small distances, this overlap becomes so pronounced that the exciton binding energy drops below that of 3D Cs$_2$AgBiBr$_6$, where Br atoms are shared rather than stacked. This effect is likely an artifact of fixed atomic positions in our simulations; in reality, octahedral tilting would reduce Br–Br overlap. These trends agree with recent results on Pb-based 2D perovskites with $n=1$ \cite{chen2023tunable}, confirming that excitons in both systems are shaped by similar quantum and dielectric confinement effects, despite their differing binding strengths and spatial extent.

In conclusion, we have used \emph{ab initio} $GW$+BSE calculations of experimental and model structures to investigate the effects of structural distortions on the optoelectronic properties of 2D Ag/Bi double perovskites. These distortions of Ag-Br bonds are driven by dimensional confinement and lead to a different direct band gap location for the experimentally synthesized materials with $n=1$ and $n=2$ perovskite layers. Because of their different orbital characters, the electronic bands at these two points form different exciton states. This work provides a crucial interpretation of the unexpected trend in exciton binding energy in these systems, highlighting the important role of structural distortions in particular in 2D materials where directional bonding interactions determine the energy and symmetry of interband transitions. We expect similar effects in other 2D perovskites in which metal $d$ orbitals contribute to the band edges, such as 2D Ag/Sb, Ag/In, and Ag/Fe double perovskites \cite{connorUnderstandingEvolutionDouble2023}. Furthermore, we believe that the relationship between Ag-halide bond distortions and exciton self-trapping which has recently been reported for several 3D double perovskites \cite{Luo2018, baskurtChargeLocalizationCs2AgBiBr62023} warrants further investigation. Finally, our systematic calculations also shed light on the evolution of exciton binding energies with respect to changes in stacking patterns and interlayer distances.

The authors thank Marina Filip for discussions and her review of this manuscript. The authors acknowledge funding by the Dutch Research Council (NWO) through Grant Nos. OCENW.M20.337 and VI.Vidi.223.072. We acknowledge computing resources provided by the Dutch national supercomputing center Snellius supported by the SURF Cooperative, PRACE and EuroHPC for awarding access to the Marconi100 and Leonardo Booster supercomputers at CINECA, Italy. This research used resources of the Oak Ridge Leadership Computing Facility, which is a U.S. Department of Energy Office (U.S. DOE) Science User Facility supported under Contract DE-AC05-00OR22725 (accessed through the INCITE and SummitPlus programs).

\section*{Author Contributions}
P.L. contributed to formal analysis, investigation, data curation, visualization and writing -- original draft. R-I. B. contributed to formal analysis, investigation, visualization and data curation. L.L contributed to conceptualization, funding acquisition, project administration, supervision, visualization and writing -- original draft and review \& editing. 



%
\end{document}